\documentclass[a4paper,11pt]{article}
\pdfoutput=1 
\usepackage{jheppub} 

\usepackage[T1]{fontenc} 
\usepackage{caption} 
\usepackage{subcaption}

\usepackage[latin9]{inputenc}
\usepackage{float}
\usepackage{amsmath}
\usepackage{amssymb}
\usepackage{graphicx}
\usepackage{esint}
 \usepackage{hyperref}
\usepackage{comment}
\usepackage{color}
\usepackage{microtype}
\usepackage{cleveref}
\usepackage{breakurl}
\usepackage{bbm}
\newcommand{\be}{\begin{equation}}
\newcommand{\ee}{\end{equation}}
\newcommand{\ben}{\begin{displaymath}}
\newcommand{\een}{\end{displaymath}}
\newcommand{\bea}{\begin{eqnarray}}
\newcommand{\eea}{\end{eqnarray}}
\def\K{K{\"a}hler}
   \newcommand{\rf}[1]{(\ref{#1})}

\def\be{\begin{equation}}
\def\ee{\end{equation}}
\def\bea{\begin{eqnarray}}
\def\eea{\end{eqnarray}}
\def\ba{\begin{array}}
\def\ea{\end{array}}
\def\bit{\begin{itemize}}
\def\eit{\end{itemize}}

\newcommand{\N}{\mathcal{N}}
\newcommand{\cN}{\mathcal{N}}

 \makeatletter

\allowdisplaybreaks

\makeatother

\makeatletter
\DeclareRobustCommand{\rcite}[1]{%
  \rcite@aux#1,\@nil{#1}%
}
\def\rcite@aux#1,#2\@nil#3{%
  \if\relax#2\relax
    Ref.~\cite{#3}%
  \else
    Refs.~\cite{#3}%
  \fi
}
\makeatother

\hypersetup{
    colorlinks = true,
    citecolor = {blue},
    linkcolor = {blue},
    urlcolor = {blue},
}

 \title{\rm {\bf \huge   Mass Production of  Type IIA dS Vacua}}

\author{Renata Kallosh}
\author{and Andrei  Linde}
\affiliation{Stanford Institute for Theoretical Physics and Department of Physics, Stanford University, Stanford, CA 94305, USA}
\emailAdd{kallosh@stanford.edu}
\emailAdd{alinde@stanford.edu}

\notoc

\abstract{A three-step procedure is proposed in type IIA string theory to stabilize multiple moduli in a dS vacuum. 
The first step is to construct a progenitor model with a localized stable supersymmetric Minkowski vacuum, or a discrete set of such vacua.   It can be done, for example, using two non-perturbative exponents in the  superpotential for each modulus, as in the  KL  model \cite{Kallosh:2004yh}.  A large set of supersymmetric Minkowski vacua with strongly stabilized moduli is protected by a theorem on stability of these  vacua  in absence of  flat directions \cite{BlancoPillado:2005fn}.  The second step involves a parametrically  small downshift to a supersymmetric AdS vacuum, which can be achieved by a small  change of the superpotential.  The third step is an uplift to a dS vacuum with a  positive cosmological constant using the $\overline {D6}$-brane contribution  \cite{Kallosh:2018nrk,Cribiori:2019bfx}. Stability of the resulting dS vacuum is inherited from the stability of the original  supersymmetric Minkowski vacuum if the supersymmetry breaking in dS vacuum is parametrically small  \cite{BlancoPillado:2005fn,Kallosh:2011qk}.  
}

\begin{document}

\maketitle

   \newpage

 \tableofcontents{}

 \parskip 7.5pt 

\section{Introduction}

The problem of construction of de Sitter vacua from string theory compactified to four dimensions is important and complicated. Some  of the solutions of this problem include the KKLT construction in  type IIB superstring theory   \cite{Kachru:2003aw}, a 
large volume compactification  in \cite{Balasubramanian:2005zx}, and  the  earlier constructions of dS vacua   in a non-critical string theory in \cite{Maloney:2002rr}. The KKLT construction, in the effective 4d supergravity approach,  involves a two-step procedure. First, using stringy perturbative and non-perturbative contributions to an effective superpotential, one can stabilize the volume of extra dimensions in a supersymmetric AdS vacuum. Secondly, with the help of an anti-D3-brane, one can uplift this anti-de Sitter vacuum with a negative cosmological constant to a de Sitter vacuum with a positive cosmological constant. In four-dimensional $\N=1$ supergravity, this second step  can be  described by a non-linearly realized supersymmetry and a nilpotent multiplet \cite{Ferrara:2014kva, Kallosh:2014wsa, Bergshoeff:2015jxa, Kallosh:2015nia, Garcia-Etxebarria:2015lif, Dasgupta:2016prs, Vercnocke:2016fbt, Kallosh:2016aep, Bandos:2016xyu, Aalsma:2017ulu, GarciadelMoral:2017vnz, Cribiori:2017laj, Aalsma:2018pll, Cribiori:2018dlc, Cribiori:2019hod}. A detailed derivation of the KKLT construction of de Sitter vacua from ten dimensions was presented recently in \cite{Hamada:2018qef,Hamada:2019ack,Kachru:2019dvo}. 

Meanwhile for many years the situation in the  type IIA string theory remained unsatisfactory, as illustrated by numerous no-go theorems, see e.g.  \cite{Kallosh:2006fm, Hertzberg:2007wc, Haque:2008jz, Flauger:2008ad, Danielsson:2009ff, Wrase:2010ew, Shiu:2011zt, Junghans:2016uvg, Andriot:2016xvq, Junghans:2016abx, Andriot:2017jhf}.   However, recently a possibility of obtaining metastable de Sitter vacua in type IIA supergravity was proposed in \cite{Kallosh:2018nrk}. It was explained there that by adding pseudo-calibrated anti-Dp-branes wrapped on supersymmetric cycles, and, more specifically,  anti-D6-branes, one can generalize the effective four-dimensional supergravity derived from string theory in a way that it includes a nilpotent multiplet. 
An explicit  KKLT-like two-step construction in the type IIA string theory has  been presented in \cite{Cribiori:2019bfx}. It involves KKLT-type nonperturbative superpotentials for each of the moduli. The existence  of a stable dS state  in the STU model presented in \cite{Cribiori:2019bfx} was confirmed there by a direct numerical analysis. Thus one may argue that the situation with dS vacua  in the  type IIA string theory  became   similar to the situation in  the type IIB case.

In this paper we would like to suggest a procedure which allows  mass production of dS vacua in the  type IIA string theory based on models which have as their progenitors stable supersymmetric Minkowski vacua.   The existence of  supersymmetric Minkowski vacua depends on existence of some points  $z_a^0$, where both the superpotential  and its covariant derivatives vanish,
  \be
W(z^a)=0\, ,  \qquad D_a W=\partial_a W =0, \qquad   {\rm at } \ \ \qquad z^a= z^a_0 \ .
\label{proj} \ee 
We additionally require the absence of flat directions. Such vacua are  stable, and they remain stable under sufficiently small deformations of the original theory. 

In particular, these vacua remain stable after a small downshift to AdS and a small uplift to dS.
The main idea of the scenario of  mass production of dS vacua in  type IIA supergravity is to start with a  set of stable Minkowski vacua, which, as we will see, can be relatively easy to find. Then one should proceed with a small downshift   to a larger set of the  stable AdS vacua,  and  consequently uplift them to an even larger set of stable dS vacua. As we will show, these vacua retain stability of the original supersymmetric Minkowski vacua if supersymmetry breaking in dS vacua is sufficiently small.

A single-field realization of this mechanism,  called KL model, was originally developed as a specific version of the KKLT construction in type IIB string theory   \cite{Kallosh:2004yh}. 
The basic idea of the KL scenario is that instead of searching for a stable AdS vacuum, one finds a supersymmetric Minkowski vacuum using racetrack superpotentials, and then uplift these potentials to dS.\footnote{In the KKLT type single-exponent models, Minkowski solutions of equations \rf{proj} are only possible at infinity of the moduli space, when the exponent vanishes.  In the two-exponent KL-type models,   the potential has a supersymmetric Minkowski minimum at  finite values of the moduli, for a broad range of the parameters.}
 The resulting potentials  have  a set of isolated minima without flat directions, with high barriers separating them, and with moduli masses that can be very large   \cite{BlancoPillado:2005fn,Kallosh:2011qk}.   Most of the properties of these potentials, including the positions of their minima, the masses of the moduli at the minimum, etc., can be found analytically for a broad range of parameters.

 A particular example of the scenario  of  mass production of stable dS vacua developed in this paper is based on a generalization of the KL scenario to the problem of moduli stabilization in the type IIA string theory. While many of our results are model-independent, a significant simplification occurs in the models where  the superpotential of string theory moduli consists of a sum of KL-type superpotentials for each of the moduli.   In this case we can find a stable supersymmetric Minkowski vacuum state by a simple analytical calculation, which yields a strictly positive definite mass matrix for all string theory moduli, without flat directions. These results remain practically unaffected and ensure stability of the uplifted dS if the resulting SUSY breaking in the uplifted dS state is not too large. This is the special property of the KL construction and the of set of more general models discussed in this paper. In the more conventional version of the KKLT mechanism, the  barrier protecting the vacuum state typically disappears in the limit of small SUSY breaking.

Below we give a short description of the three stages we will use here to build a stable dS vacuum state in supergravity in general, and in particular in supergravity models of the KL type, inspired by type IIA string theory.
The first stage of our proposal it to find a general class of models with stable Minkowski minima without flat directions.
The basic property of all Minkowski vacua \rf{proj} with $V_{Mink}=0$    in a theory with multiple chiral multiplets is that  the holomorphic-anti-holomorphic mass matrix  of scalars is a square of the fermion mass matrix, 
\be
(V_{a}{} ^{b})_{Mink} = m_{a c} \bar m^{c  b} \ .
\label{square}\ee
Here $V_{a}{} ^{b}=  D_{a} D_{ \bar b}  V \, g^{ b \bar b} $, and  the chiral fermion mass term in the action is
$-{1\over 2} {\it m}_{ab} \chi^a \chi^b +hc$, see for example \cite{Freedman:2012zz}.  
The second derivatives of the potential over scalars form a strictly positive definite matrix under a condition that all fermion mass eigenvalues are non-vanishing and the matrix $V_{a}{} ^{b} $ is not degenerate. Thus, the first goal of our proposal is to find a class of supergravities where such properties are realized, 
\be
(V_{a}{} ^{b})_{Mink} = m_{a c} \bar m^{c  b} >0\ .
\label{square1}\ee
This was explained in  \cite{BlancoPillado:2005fn} where also the importance of the absence of flat directions of the potential  was discussed, and some examples of local KL-type Minkowski minima without flat directions were presented.

In this paper we will construct a  broad class of multi-field  models which satisfy all required conditions mentioned above. This is the basis of our  work, but two further steps are still required. 

The second stage of our proposal is the downshift to a supersymmetric AdS. A  change of the  superpotential by adding to it  a small term $\Delta W$   transforms the original Minkowski vacuum into a supersymmetric  AdS vacuum with    $V_{AdS}=  - 3 e^K | W|^2$
where $e^{K}   | W|^2 = m_{3/2}^2 $ and $W\approx  \Delta W$, up to higher orders in $|\Delta W|$.

  In section \ref{mass} we will show that the downshifted AdS state remains a minimum of the potential under the condition that the scale of the smallest (non-vanishing) eigenvalue $m_\chi$ of the fermion matrix $m_{a c}$ is much greater than the scale of the gravitino mass
\be
m_\chi \gg m_{3/2} \ ,
\label{cond1}\ee
Constraints of the type of \rf{cond1} are known as conditions of strong moduli stabilization \cite{Kallosh:2004yh,BlancoPillado:2005fn,Kallosh:2011qk,Linde:2011ja,Dudas:2012wi,Kallosh:2014oja}. Note that  the mass matrix of the strongly stabilized Minkowski minimum practically does not depend on $\Delta W$ if the term $\Delta W $ is small enough. Thus for sufficiently small $\Delta W$ (and, consequently, $m_{3/2}$) the stable Minkowski vacuum remains a stable AdS after the downshift. In this way we construct AdS model with many different values of the gravitino mass.

Finally, the third stage of our proposal is the uplift to dS, using the anti-D6-brane, or an equivalent nilpotent multiplet $X$ in supergravity  \cite{Kallosh:2018nrk,Cribiori:2019bfx}. The dS  potential 
 has a minimum at the position of moduli slightly shifted from the ones they had in AdS. The condition for dS state to be a minimum requires that the total supersymmetry breaking, including a nilpotent direction,   $F^2= e^K |DW|^2$ is small, in addition to \rf{cond1} so that
 \be
|F|^2\ll  m_\chi ^2 \ ,   \qquad   m_{3/2}^2 \ll m_\chi ^2   \ ,
\label{smallerF}
\ee
and the dS potential is positive if 
\be
\Lambda = |F|^2 - 3 m_{3/2}^2 >0 \ .
\label{cc}\ee
A general analysis  of stability of the uplifted dS vacuum, based on conditions \rf{smallerF}, \rf{cc} is performed in multifield KL models in Secs. \ref{multi}, \ref{STUsect} and in section \ref{mass} for general choice of $K$ and $W$.  In section \ref{mass}  the result is derived using an explicit expression for the second derivatives of the potentials in supergravity.
 If the required  conditions are met (and they are always met for sufficiently small $\Delta W$ and $F$), the uplifted dS vacua inherit their stability property from their Minkowski progenitor.  Note that this does not mean that the supersymmetry breaking must occur on a very small scale, such as  the electroweak scale which is $10^{-15}$ in Planck units. We only require that the susy breaking should be smaller than the typical scale involved in the supersymmetric Minkowski vacuum construction, which can be extremely high in supergravity. 
 
 One may wonder why would we need to start with a supersymmetric Minkowski vacuum and then downshift to AdS, instead of starting directly with AdS, as in the standard version of the KKLT construction? And why do we want to downshift from the Minkowski vacuum before uplifting to dS, whereas one can make the uplift directly from the Minkowski vacuum?
 
 The reason we want to start with Minkowski rather than AdS is that the discrete supersymmetric Minkowski state is always a minimum and can be strongly stabilized. This feature continue protecting stability of the AdS and dS vacua originating from the supersymmetric Minkowski state.  The downshift is required because without it, the vacuum with the observable value of the cosmological constant uplifted from the Minkowski state with $V=0$ to the dS state with $V_{dS} = 10^{{-120}}$ would have $F^{2} \sim 10^{{-120}}$ and therefore the  SUSY breaking would be too small. A controllable downshift   disentangles SUSY breaking form the smallness of the cosmological constant. 
 
 But if a  general goal is to study a possibility to construct stable dS vacua with different values of $V_{dS}$, one can certainly obtain many of them by an uplift directly from the supersymmetric Minkowski vacuum. We present an example of such an uplift without a downshift in  appendix A.

Thus, in the mass production scenario one can construct  a large variety of supersymmetric Minkowski vacua, an  even greater variety of stable supersymmetric downshifted AdS states, depending on many different ways of the downshifting, and an even greater number of stable dS vacua obtained either after the downshift, or by a direct uplift from one of the many possible supersymmetric Minkowski vacua.  The three-step procedure allows  to rely on  multiple  AdS vacua and many versions of uplifts. This may help to address the smallness of the cosmological constant $\Lambda\approx 10^{-120}$ using the string theory landscape scenario.

 Our investigation will be mostly analytical, but in appendix A we will present results of a numerical analysis of a particular string theory motivated model illustrating our main conclusions.

\section{A single field scenario}

\subsection{A general theory}

Consider first a theory  of a single field $T  = t +i \, \theta $ with 
\be
K= -\ 3 \ln(T +\overline{T }) \ ,
\qquad 
W(T) = W_{0} + W_{1}(T) \ .
\ee
In this theory one has
\be
DW = -3{W_{0} + W_{1}(T)\over T+\bar T} +{dW_{1}(T)\over dT} \ .
\ee
This theory has a supersymmetric Minkowski vacuum at some value of the field $T = T_{0}$ if
\be\label{zerow}
W = W_{0} + W_{1}(T_{0}) = 0
\ee
and
\be\label{zerodw}
DW = -{3W_{0} + W_{1}(T_{0}) \over T_{0}+\bar T_{0}} +\left({dW_{1}(T)\over dT}\right)_{T=T_{0}} =0\ .
\ee
Using \rf{zerow}, one can represent \rf{zerodw} as a simple condition
\be \label{zeroder}
W' |_{T=T_0}  =0 .
\ee
Our strategy of finding a theory with a supersymmetric Minkowski vacuum state with a potential having a localized minimum consists of the following steps. First of all, one should find a function $W(T)$ such that equation \rf{zeroder} for its derivative is satisfied at a localized point $T_{0}$, or at a  series of points, rather than along a flat direction. Then one should take $W_{0}= -W(T_{0})$.
For example, one may consider a superpotential
\be
W(T) =(T-T_{0})^{2} \ .
\ee
which satisfies the required conditions
\be
W(T_0)=W'(T_{0})=0 \ .
\ee
This corresponds to a supersymmetric stable Minkowski vacuum state. 

For simplicity, we will consider the theories where $T_{0} =  t_{0} +i \, \theta_{0}$ is real, i.e. $T_{0} =  t_{0}$. The mass of the field $T$    at the minimum of the potential is given by
\be
m^{2} = {2t_{0}^{2}\over 3}\, {d^{2}V\over dt^{2}}
 = {2\,t_{0}\over 9} \,   \bigl(W''(t_{0})\bigr)^{2} \ .
\ee
Here all derivatives are taken at $T =   t_{0}$. An extra coefficient ${2t_{0}^{2}\over 3}$ relating ${d^{2}V\over dt^{2}}$ to $m^{2}$ appears due to the fact that  $T$ is not canonically normalized. 

If one adds a small constant $\Delta W$ to the superpotential, the Minkowski vacuum state with $V(t_{0}) = 0$ becomes a stable supersymmetric AdS vacuum, with   $DW= 0$   \cite{Kallosh:2011qk}. For sufficiently small  $\Delta W$, one finds  that the shift of the field from $t_{0}$ is negligibly  small, and the leading contribution to $W$ is given by $\Delta W$. As a result, the vacuum energy becomes 
\be
V_{\rm AdS} = -3e^{K}(\Delta W)^2 \approx -{3(\Delta W)^2\over 8 t_{0}^{3}} \ .
\label{vads0}
\ee 
An uplifting of this vacuum state to a dS state with a nearly vanishing $V_{\rm AdS}$ can be achieved due to the anti-D3-brane contribution represented by a nilpotent field $X$. One can describe it by adding a term $\mu^2 X$ to the superpotential and a term $X\bar X$ to the Kahler potential, and then taking $X = 0$ after calculating all quantities of interest,  such as the potential. One finds that the uplift of the state with $V_{AdS}$ \rf{vads0} to a dS state with a nearly vanishing cosmological constant $V_{dS} \sim 10^{{-120}}$   requires the uplift parameter 
\be
\mu^{4} =  3\, (\Delta W)^2 \ .
\ee
The gravitino mass after uplifting is given by 
\be
m^{2}_{3/2} =  e^{K}(\Delta W)^2 = { (\Delta W)^2\over 8 t_{0}^{3}} ={|V_{AdS}|\over 3}= { \mu^4\over 24 t_{0}^{3}} \ .
\ee
Once again, for sufficiently small values of $\Delta W$, the required uplifting does not affect the strong stabilization of the original Minkowski vacuum state. An approximate criterion  of applicability of these conclusions and the estimates is the requirement that the gravitino mass should be much smaller than other mass parameters describing the potential, see a detailed discussion of this issue in Sect. \ref{mass}. 

\subsection{The basic KL model}\label{baseKL}

The simplest string theory related realization of the general mechanism described above, proposed in~\cite{Kallosh:2004yh}, is to use the \K\ potential
\be 
K = -3 \ln (T+\bar T) + X\bar X, 
\ee
and the
racetrack superpotential  with two exponents 
\be
W_\text{KL}(T,X)  =W_{0} +Ae^{-aT}- Be^{-bT}  +   \mu^2 X\ ,
\label{adssup}
\ee
 which can arise, for example, in the presence of two stacks of D7 branes wrapping homologous 4-cycles. Gaugino condensation on the first one is responsible for the KKLT-type term $Ae^{-aT}$, the second one for the term $- Be^{-bT}$. If there are $N_{1}$ branes in the first stack, and $N_{2}$ branes in the second one, one has $a = {2\pi/  N_{1}}$ and $b = {2\pi/  N_{2}}$. 
The parameters $A$ and $B$  depend on  the  values at which the complex structure moduli are stabilized \cite{Burgess:1998jh,Baumann:2006th,Baumann:2010sx}, and therefore one may expect $A$ and $B$ to span large range of possible values, 
due to the  large variety of vacua in the string theory landscape.

In what follows we will consider the models  where $a, b, A, B > 0$,   and 
\be\label{w0stab}
W_0=  -A \left({a\,A\over
b\,B}\right)^{a\over b-a} +B \left ({a\,A\over b\,B}\right) ^{b\over b-a}  \ .
\ee

As a first step of our analysis, we look for supersymmetric Minkowski minima, which can be found by solving   equations  
\be\label{susy} 
W(t_{0})=0 \ , \qquad DW(T_{0})=0 \ ,
\ee
for $\Delta W = 0$, $\mu^2 = 0$.   Looking for  the solutions in the full complex plane $T = t+ i\, \theta$, one finds
\be\label{fullcompl}
T_{0} = t_{0} + i \theta_{0}={1\over a-b}\left(\ln  {a\,A\over b\,B}  +i\,  {2n \pi } \right)\ .
\ee
The last term reflects the fact that the potential $V(T)$ is periodic in the ${\rm Im\ T}$ direction with a period ${2\pi\over a-b }$. It has a series of  minima with $V = 0$ shown in \rf{fullcompl}, separated from each other by high barriers. Importantly, the potential does not have any flat directions that could be destabilized after small deformations of the potential. Without  loss of generality, it is sufficient to limit our investigation to the supersymmetric Minkowski vacuum with $\theta = 0$ and 
\be\label{sigmacr}
 t_{0} ={1\over a-b}\left(\ln  {a\,A\over b\,B}  \right)\ .
\ee
The mass squared of both real and imaginary components of the field $T$ at the supersymmetric Minkowski minimum with $V = 0$   is given by
\be
m^{2}_{t_{0}} =  {2t_{0}\over 9} \bigl(W''(t_{0})\bigr)^{2} = {2\over 9} a A b B  (a -b )\left(a A \over b B  \right)^{-{a +b \over a -b }}  \ln {a\, A\over b\, B}\ .
\ee

 Adding a small correction $\Delta W$ to $W_{0}$ makes this minimum AdS, with the value of the potential at its minimum downshifted from $V = 0$ to
\be
V_{\rm AdS}  = -{3(\Delta W)^2\over 8 t_{0}^{3}}= - {3\over 8} \left({a - b \over  \ln \left({a A\over b B}\right)}\right)^{3}\, (\Delta W)^2 \ .
\label{vads}
\ee 
The supersymmetric vacuum \rf{sigmacr} exists for any values of $a, A, b, B$ satisfying the condition $a > b$ and $aA > bB$. 

 It is useful to understand the properties of the KL potential and the mass of the volume modulus $m^{2}_{t_{0}}$ under the simultaneous  rescaling of the parameters $A \to C A$, $B\to CB$. This rescaling does not affect the position of the minimum $t_{0}$, but  it increases the value of $W_{0}$ and the mass of the field $T$  by a factor $C$, and it increases  the height of the barrier stabilizing the minimum in the KL potential at $ t_{0}$ by a factor $C^2$. 
Meanwhile the simultaneous rescaling $a \to c a$ and $b \to c b$  decreases $t_{0}$ by a factor of $c$, increases $m^{2}_{t_{0}}$ by a factor of  $c^{{3}}$, and increases the height of the barrier by a factor of $c$ \cite{Kallosh:2011qk}. Thus one has a significant freedom to change the properties of the KL potential.

This minimum  can be easily uplifted to dS while remaining strongly stabilized  by taking $\mu^{4} >  3\, (\Delta W)^2$ \cite{Kallosh:2004yh,BlancoPillado:2005fn,Kallosh:2011qk,Linde:2011ja,Dudas:2012wi,Kallosh:2014oja}. Importantly, the height of the barrier in this scenario is not related to supersymmetry breaking and can be arbitrarily high. Therefore, this  potential can be strongly stabilized by a proper choice of the parameters, which makes it  especially suitable for being a part of the inflationary theory~\cite{Kallosh:2011qk,Dudas:2012wi,Kallosh:2018zsi}.  Some aspects of implementation of this scenario in string theory are discussed in \cite{Kallosh:2019axr}.

Now we will generalize the single-field  models discussed above to the multi-field constructions capable of stabilizing many string theory moduli in AdS, Minkowski and dS spaces.

\section{Multifield supersymmetric Minkowski vacua and their dS uplift}\label{multi}

\subsection{Finding a supersymmetric Minkowski vacuum}

Consider a theory of the fields $T_{i}$,\, $i = 1,2,...,n$, with a general \K\ potential
\be\label{genK}
K=  K(T_{1},\overline{T_{1}}, ..., T_{n},\overline{T_{n}}) \ ,
\ee
and a superpotential
\be\label{verygenW}
W = W(T_{1},...,T_{n}) \ .
\ee
To find a supersymmetric Minkowski vacuum at $T_{i}= T^{0}_{i}$ one should impose the  conditions  
\be\label{GGenW0}
W(T^{0}_{1},...,T^{0}_{n})=  0 \ ,
\ee
and $D_{T_{i}}W  = 0$, which should be satisfied at $T_{i}= T^{0}_{i}$ for each $i$. For the theory  \rf{genK}, \rf{verygenW}, this last condition, in combination with \rf{GGenW0}, leads to $n$ independent equations
\be
\partial_{T_{i}}W(T_{1},...,T_{n})    = 0 \ ,
\ee
which should be satisfied at  some point  $T_{i}= T^{0}_{i}$, $i = 1, ..., n$.

In this context, the procedure of constructing a theory which has a supersymmetric Minkowski vacuum at a given point $T_{i}= T^{0}_{i}$ is very simple. First of all, one should find a function  $U(T_{1},...,T_{n})$ which has an extremum at this point. Then the theory with the superpotential 
\be\label{fullW}
W = U(T_{1},...,T_{n})-U(T^{0}_{1},...,T^{0}_{n})  
\ee
has a supersymmetric Minkowski vacuum  at  $T_{i}= T^{0}_{i}$. Note that this result is valid independently of the choice of the \K\ potential.


We will analyze a general class of such models in section \ref{mass}, but in this section and in section \ref{STUsect}  we will concentrate on  superpotentials
\be\label{genW}
W = W_{0} + \sum\limits_{i=1}^n  W_{i} (T_{i}) \ .
\ee
To find a supersymmetric Minkowski vacuum at $T_{i}= T^{0}_{i}$ one should have
\be\label{gW0}
W(T^{0}_{i})= W_{0} + \sum\limits_{i=1}^n  W_{i}(T^{0}_{i}) = 0 \ ,
\ee
and 
\be
\partial_{T_{i}}W_{i}  = 0 
\ee
at $T_{i}= T^{0}_{i}$. We will  consider the \K\ potential  
\be\label{genK1}
K= -\sum\limits_{i=1}^n  N_{i}\, \ln(T_{i}+\overline{T_{i}}) \ ,
\ee
and represent the fields $T_{i}$ as $T_{i}= t_{i}+ i\theta_{i}$. The supersymmetric vacua that we are going to explore,  as well as the vacua after the downshifting and uplifting, are positioned at $t_{i}\not = 0$, and $\theta_{i} = 0$. 
A useful general  result simplifying our investigation is that in the models \rf{genK1}, \rf{genW}  in the minima with $\theta_{i} = 0$ one has
\be
{dV\over d\theta_{i}} = 0\ , \qquad {d^{2}V\over dt_{i} d\theta_{j}} = 0 \ .
\ee 
The only non-zero second derivatives of the potential are ${d^{2}V\over dt_{i} dt_{j}}$ and  ${d^{2}V\over d\theta_{i} d\theta_{j}}$. In other words, the corresponding matrix is  block-diagonal. This simplifies investigation of stability of the vacuum state by reducing it to the analysis of stability with respect to $t_{i}$, and to a  separate analysis with respect to $\theta_{i}$. 

Moreover, at the supersymmetric Minkowski vacuum, the matrices ${d^{2}V\over dt_{i} dt_{j}}$ and ${d^{2}V\over d\theta_{i} d\theta_{j}}$, as well as the mass matrix of the canonically normalized fields at the minimum of the potential, are diagonal. The masses of the fields of the canonically normalized counterparts of the fields $t_{i}$ and $\theta_{i}$ are equal to each other, and are given by 
\be
 m^{2}_{i} = {2t^{2}_{i}\over N_{i}}{d^{2}V\over dt_{i} dt_{j}} = {2t^{2}_{i}\over N_{i}}{d^{2}V\over d\theta_{i} d\theta_{j}}  \ .
\ee
The final result is
\be\label{mmatr}
 m^{2}_{i} =   2^{4}\,  \prod_{j=1}^{n} (2t_{j})^{-N_{j}}\,  \left({t_{i} ^{2}\, \over N_{i}} {d^{2}W\over d{t_{i}}^{2}}\right)^{2} \ ,
 \ee
where the whole expression should be evaluated at the Minkowski minimum $t_{i} = t^{0}_{i}$, $\theta_{i} = 0$.

Adding a small term $\Delta W$ to the superpotential does not affect the vacuum stability and makes only a small contribution to the mass matrix \rf{mmatr}, but it shifts the minimum to the AdS state with the negative cosmological constant
\be
V_{\rm AdS} \approx - 3\, (\Delta W)^2 \prod_{j=1}^{n} (2t^{0}_{j})^{-N_{j}} \ .
\label{vadsmany}
\ee

As we will see in the example of the STU model in section \ref{STUsect}, the uplift of the AdS vacuum can be achieved by adding to the superpotential a term $\mu^{2} X$, and by adding to the \K\ potential an uplifting term, which can be generally represented as 
\be
\Delta K = {X \bar X\over (T_{1} + \overline{T_{1}})^{k_{1}}\cdot ... \cdot (T_{n} + \overline{T_{n}})^{k_{n}}} \ .
\ee
A subsequent addition of the nilpotent field contribution $\mu^2 X$ uplifts the vacuum state to a stable dS state. For small $\Delta W$ and $\mu$, we have the value of the potential at the minimum
\be
V_{\rm min} =   \mu^{4 }\prod_{i=1}^{n} (2t_{i})^{k_{i}-N_{i}}-3\, (\Delta W)^2 \prod_{i=1}^{n} (2t^{0}_{i})^{-N_{i}}  \ .
\ee
This corresponds to a stable dS state for 
\be\label{stabledS}
\mu^{4 } > 3\, (\Delta W)^2 \prod_{i=1}^{n} (2t^{0}_{i})^{-k_{i}} \ .
\ee
This state is a  dS vacuum with an extremely small cosmological constant for 
\be\label{smalldS}
\mu^{4 } \approx 3\, (\Delta W)^2 \prod_{i=1}^{n} (2t^{0}_{i})^{-k_{i}} \ .
\ee

The gravitino mass after uplifting to a (nearly Minkowski) dS vacuum state is given by 
\be
m^{2}_{3/2} =  e^{K}(\Delta W)^2 = (\Delta W)^2 \prod_{j=1}^{n} (2t^{0}_{j})^{-N_{j}} ={|V_{AdS}|\over 3} \ .
\ee

\subsection{Multifield KL scenario}

As a particular string theory motivated example, one can take a theory with the racetrack superpotential used in the KL scenario  for each moduli:
\be\label{KLsup}
W = W_{0}   +\sum\limits_{i=1}^n  A_{i} e^{-a_{i}T_{i}} -  \sum\limits_{i=1}^n  B_{i} e^{-b_{i}T_{i}}  \ , \qquad K= -\sum\limits_{i=1}^n  N_{i}\, \ln(T_{i}+\overline{T_{i}}) \ .
\ee
Using the prescription outlined above in application to the models with either of these two superpotentials one finds a stable supersymmetric Minkowski vacuum without any flat directions for all moduli. All  fields $T_{i}$ are fixed in the supersymmetric Minkowski vacuum state at their values satisfying the equation
\be
a_{i}A_{i}e^{-a_{i}T^{0}_{i}} = b_{i}B_{i}e^{-b_{i}T^{0}_{i}}
\ee
for each $i$.  The real field solutions, for each $i$, are $T^{0}_{i} = t^{0}_{i}$
\be\label{mingen} 
t^{0}_{i}= {1\over {a_{i}-b_{i}}}\, \ln{a_{i} A_{i}\over b_{i}\,B_{i}} \ .
\ee
This finally yields, for both choices of the superpotential,
\be\label{w0}
W_{0}= -\sum_{i}  A_{i} e^{-a_{i}t^{0}_{i}}+ \sum_{i}  B_{i} e^{-b_{i}t^{0}_{i}} \ .
\ee
The  \K\, potential and superpotential of the full  theory involve a nilpotent multiplet $X$
\be
K= -\sum\limits_{i=1}^n  N_{i}\, \ln(T_{i}+\overline{T_{i}}) + K_{X\bar X} (T_i, \bar T_i) X \bar X \ ,
 \ee
\be\label{fullsuper}
W =  \sum\limits_{i=1}^n A_{i} e^{-a_{i}T_{i}} -  \sum\limits_{i=1}^n  B_{i} e^{-b_{i}T_{i}}- \sum\limits_{i=1}^n  A_{i} e^{-a_{i}t^{0}_{i}}+  \sum\limits_{i=1}^n  B_{i} e^{-b_{i}t^{0}_{i}} +\Delta W +\mu^{2}\, X\ .
\ee
This model  has a supersymmetric Minkowski vacuum at  $T_{i}= t^{0}_{i}$ for $\Delta W=0$, $\mu^2=0$
 The masses of all moduli in the Minkowski vacuum, or in an AdS or dS vacuum with small $\Delta W$ and $\mu^2$, are given by \rf{mmatr},
where one should use equation \rf{mingen} for $t^{0}_{i}$, and, for each $i$,
\be\label{dder}
\left({d^{2}W\over d{t_{i}}^{2}}\right)^{2} = a_{i}A_{i}b_{i}B_{i} (a_{i}-b_{i})^{2} \left(a_{i}A_{i}\over b_{i}B_{i} \right)^{-{a_{i}+b_{i}\over a_{i}-b_{i}}} \ .
\ee
In section \ref{baseKL} we mentioned that the   large freedom of choice of the parameters  $a_{i}$, $A_{i} $, $b_{i}$, $B_{i} $ gives us a possibility to significantly modify the properties of the potential.

If one takes into account the tiny correction $\Delta W$, the minimum of the potential shifts to the AdS state with $V_{AdS}$ proportional to $-(\Delta W)^{2}$ as shown in \rf{vadsmany}. Note that this general result does not depend on the choice of the superpotentials $W_{i} (T_{i})$. 

With an account taken of the small  contribution represented by a nilpotent superfield $X$, the minimum can be easily uplifted to dS. The uplifted vacuum describes a stable dS state for sufficiently small $\Delta W$ and $\mu$ under the conditions \rf{stabledS}, \rf{smalldS}.  A general discussion of  stability conditions can be found in   \cite{Kallosh:2004yh,BlancoPillado:2005fn,Kallosh:2011qk} and in section \ref{mass} of our paper.

\section{An example: STU model}\label{STUsect}

As an example particularly relevant for the dS construction in the type IIA string theory, we will consider here the STU model, extending the analysis of  \cite{Kallosh:2018nrk,Cribiori:2019bfx}. The KL generalization of the STU model considered in \cite{Cribiori:2019bfx} can be represented, in slightly different notations, as follows:
\be
W = W_{0}+\Delta W+ \sum\limits_{i=1}^3  W_{i} (T_{i})+ \mu^2\, X \ ,
\ee
\be 
K = - \ln(T_{1}+\overline{T_{1}}) - 3\ln(T_{2}+\overline{T_{2}}) - \ln\Bigl((T_{3}+\overline{T_{3}})^3 -  \frac{X \bar{X}}{(T_{1}+\overline{T_{1}}) +g (T_{2}+\overline{T_{2}})}\Bigr) . 
\ee
Here $T_{1}$ stands for the field $S$, $T_{2}$ stands for $T$, and $T_{3}$ corresponds to $U$, whereas $W_{0} = f6$ in notations of  \cite{Cribiori:2019bfx}. The term $\frac{X \bar{X}}{(T_{1}+\overline{T_{1}}) +g (T_{2}+\overline{T_{2}})}$ under the logarithm represents the uplifting contribution of the $\overline{D6}$ brane, $g$ is some constant. Since $X$ is a nilpotent field, such that $X^{2}= 0$, the \K\ potential can be equivalently represented as 
\be 
K = - \ln(T_{1}+\overline{T_{1}}) - 3\ln(T_{2}+\overline{T_{2}}) - 3\ln(T_{3}+\overline{T_{3}}) +  \frac{X \bar{X}}{(T_{3}+\overline{T_{3}})^{3}((T_{1}+\overline{T_{1}}) +g (T_{2}+\overline{T_{2}}))}   \ .
\ee
The last term in this expression is responsible for the uplifting contribution to the potential, which can be represented as 
\be
V_{\overline{D6}} = {p^{2}\over t_{2}^{2}}+ {q^{2}\over t_{1}t_{2}} \ .
\ee
where both parameters $p^{2}$ and $q^{2}$ are proportional to $\mu^{2}$ and vanish in the absence of the uplifting.

Thus, this model belongs to the class of the models studied in the previous section. In particular,  for $\Delta W = 0$ and $\mu=0$, as well as for sufficiently small values of these parameters, the masses of all moduli in the Minkowski vacuum, or in a slightly uplifted dS vacuum, are given by  
\be\label{STUmassmatr}
 m^{2}_{i} =  {1\over 8 t_{1}\, t_{2}^{3}\   t_{3} ^{3}}\  \left({t_{i} ^{2}\, \over N_{i} }{d^{2}W\over d{t_{i}}^{2}}\right)^{2} \ ,
 \ee
where $N_{1}=1$ for the $S$ field, and $N_{2}= N_{3} =3$ for $T$ and $U$ fields. This expression should be evaluated at the Minkowski vacuum with $t_{i}= t^{0}_{i}$.

A particular string theory motivated set of superpotentials for the STU model is given by the KL superpotentials
\be
W_{i}(T_{i})= A_{i} e^{-a_{i}T_{i}} - B_{i} e^{-b_{i}T_{i}}  \ .
\ee
In this set of models, the mass matrix in the supersymmetric Minkowski vacuum,  as well as in its stable AdS and dS descendants, is given by \rf{STUmassmatr}, where one should take $t^{0}_{i}$ from equation \rf{mingen}, and ${d^{2}W\over d{t_{i}}^{2}}$ from \rf{dder}.

\section{Stability of dS Vacua}\label{mass}

\subsection{A Lesson from WZ Model}
The WZ model with a scalar $A$ and a pseudo-scalar $B$ and a Majorana fermion $\chi$ is given by
\bea
{\cal L}&=& -{1\over 2} (\partial_\mu  A\partial^\mu A + \partial_\mu B\partial^\mu B) -{1\over 2} M^2 (A^2+B^2) -{1\over 2} \bar \chi (\gamma^\mu \partial_\mu + M) \chi \cr
\cr
&-&{g\over \sqrt 2} \bar \chi (A+i\gamma_5 B) \chi -{M g\over \sqrt 2} (A^3+ AB^2) - {g^2\over 4}(A^2+B^2)^2 \ .
\eea
If the  non-vanishing mass of the fermion in a single chiral multiplet is $M$, unbroken supersymmetry requires that  the mass of the scalar field is also  $M$. Therefore the second derivative of the scalar potential  at the supersymmetric minimum $A=B=0$ is equal to the square of the fermion mass. 
\be
{d^2V(A, B) \over d A^2}= {d^2V(A, B) \over d B^2} = M^2 > 0 \ .
\label{WZ}\ee
This simple consequence of supersymmetry is the underlying reason of the mass production of de Sitter vacua, which we will explain below.
\subsection{Masses of Scalars and Fermions in $\cN=1$ supergravity}

Consider a general case of an arbitrary 
number of chiral matter superfields $z^{a}$, arbitrary K\"ahler potential ${\cal K}(z^{a}, \bar z^{\bar a})$, and arbitrary holomorphic superpotential $W(z^{a})$.  

We use the notation of  \cite{BlancoPillado:2005fn}, originally introduced in \cite{Kallosh:2000ve}. We use units in which $M_P=1$. In addition to the holomorphic superpotential $W(z^{a})$,
we define a {\it covariantly holomorphic superpotential} $m$, a complex gravitino mass 
\begin{equation}
e ^{ K/2}W \equiv {\it m}(z^{a}, \bar z^{\bar a})\ , 
\label{defm}
\end{equation}
which is related to the (real) gravitino mass,
\begin{equation}
  M^2_{3/2}  =|m \, \bar m| \ .
\label{defmee}
\end{equation}
The complex gravitino mass is covariantly holomorphic, which means that 
\begin{equation}
  \bar D_{\bar a}m\equiv (\partial_{\bar a} -{1\over 2} K_{\bar a} ){\it m}=0\ , \qquad D_{ a}\bar m\equiv (\partial_{ a} -{1\over 2} K_{ a} )\bar {\it m}=0 \ .
\label{cov}
\end{equation}
 We also define the \K\, covariant derivatives as
\begin{eqnarray}
D_a {\it m} =  \, \partial _a {\it m}
+{1\over 2} (\partial _a { K}) {\it m}\equiv {\it m}_a\ , \qquad \bar D_{\bar a} \bar {\it m} =  \, \partial _{\bar a}\bar {\it m}
+{1\over 2} (\partial _{\bar a}{ K}) \bar {\it m}\equiv \bar {\it m}_{\bar a} \ .
\end{eqnarray}
It means that
\be
m_a = e^{K/2} D_a W\, ,  \qquad \bar m_{\bar a }= e^{K/2} \bar D_{\bar a } \bar  W \, .
\ee
The complex masses of the chiral fermions in $\cN=1$ supergravity are equal to
\begin{equation}
  D_a D_b {\it m} \equiv {\it m}_{ab}\ , \qquad \bar D_{\bar a}\bar D_{\bar b}\bar {\it m}\equiv \bar {\it m}_{\bar a  \bar b} \ .
\label{chiral}
\end{equation}
Note that the fermion action has the matter fermion mass terms:
\be
-{1\over 2} {\it m}_{ab} \chi^a \chi^b +hc \ .
\label{chi}\ee
It might be useful to give also an alternative  expression for the fermion matrix ${\it m}_{ab}$, see for example eq. (18.16) in \cite{Freedman:2012zz}
\be
m_{ab} = e^{K/ 2} (\partial_a + K_a ) D_b W- e^{K/ 2} \Gamma^c_{ab} D_c W \ .
\label{fmass}\ee
This expression simplifies at the supersymmetric Minkowski minimum with $D_aW =W=0$
\be
m_{ab}^{Mink} = e^{K/ 2} \partial_a   \partial_b W \ .
\ee
The \K\ metric is
$g_{a\bar b}  =\partial_a \partial_{ \bar b}{K}$.
Using this notation we can rewrite the standard F-term potential $V= e^{\cal K}( |DW|^2-3|W|^2)$ as follows
\begin{equation}\label{pot}
V={\it m}_a\, g^{a\bar b} \, \bar {\it m}_{\bar b}  -3|{\it m}|^2 \equiv |{\it m}_a|^2-3|{\it m}|^2 \ .
\end{equation}
The first derivative of the potential becomes,
\begin{equation}
\partial _a V=D_a V= -2 {\it m}_a\,\bar {\it m} +{\it m}_{ab}g^{b\bar b} \,\bar {\it m}_{\bar b} \ ,
\label{diV}
\end{equation}
where we have taken into account that $ D_{ a} \bar {\it m}_{\bar b}= g_{ a \bar b} \bar {\it m}$ and that the potential is \K\, invariant.
Note that the potential has an extremum if 
\be
\partial _a V=D_a V=0 \ . 
\ee
If the extremum is supersymmetric, 
\be
{\it m}_a= \bar {\it m}_{\bar b} = 0 \ .
\ee
This condition is sufficient for the potential to have an extremum, however, the non-supersymmetric extrema are also possible if
\be
2 {\it m}_a\,\bar {\it m} ={\it m}_{ab}g^{b\bar b} \,\bar {\it m}_{\bar b} \ , \qquad {\it m}_a \neq 0\ ,  \qquad \bar {\it m}_{\bar b}\neq 0 \ .
\ee
The scalar mass matrix in supergravity at an extremum of the potential $\partial _a V=D_a V=0$ takes the following form
\begin{equation}
\mathcal{M}^2=\, \left ( \begin{array}{cc} V_{a\bar b} & V_{a b} \cr V_{\bar a \bar b} & V_{\bar a b} \cr  \end{array} \right ) .
\label{massM}
\end{equation}
\subsection{Supersymmetric Minkowski Minimum}
This case is simple and is defined by the conditions that 
\be
{\it m}={\it m}_a=0\, , \qquad V= |{\it m}_a|^2-3|{\it m}|^2 =0 \ ,
\ee
and it was studied in  \cite{BlancoPillado:2005fn}. The second derivative of the F-term potential
in N=1 supergravity is particularly simple in terms of the fermion masses of the chiral multiplets in the supersymmetric vacuum. Namely we find that
\begin{equation}
(\mathcal{M}^2_{sc})^ {Mink}=\, \left ( \begin{array}{cc} V_{a \bar b}^ {Mink}  & 0 
\cr 
\cr
0 &  V_{\bar a  b}^ {Mink}   \cr  \end{array} \right ) \ ,
\label{massMink}
\end{equation}
where
\begin{equation}
V_{a \bar b}^ {Mink} =   {\it m}_{ac}g^{c\bar c} \,\bar {\it m}_{\bar c \bar b} \ . 
\label{mix1}
\end{equation}

Thus, at the supersymmetric critical point in Minkowski space the mass matrix is block diagonal,   at each diagonal part of it we have a positive definite matrix, and consequently 
all its eigenvalues are positive.  Indeed, for an arbitrary
vector $\phi$,
\begin{equation}
\phi_a V_{a \bar{b}}^ {Mink}{\bar \phi_{\bar b}}= \phi_{a} {\it m}_{ac} \, g^{c\bar c} \,\bar {\it m}_{\bar c \bar b} \,
 \bar\phi_{\bar b} = \Phi_{c}\, g^{c\bar c} \,\bar \Phi_{\bar c}\  , 
\end{equation}
where  $\Phi_{c} = \phi_{a} {\it m}_{ac}$. The matrix $g^{c\bar c}$ is  positive definite, so we have 
$
\Phi_{c}\, g^{c\bar c} \,\bar \Phi_{\bar c} \geq 0
$
for any vector $\Phi$. Thus  ${\cal M}^2$ is  a positive 
definite matrix, namely,
\begin{equation}
\phi_a \, V_{a \bar{b}}^ {Mink} \, {\bar \phi_{\bar b}} \geq 0 \ .
\end{equation}
 for an arbitrary vector $\phi_{a}$. It  means that the second 
derivative of the potential is non-negative in all directions in the moduli space.
The conclusion reached in  \cite{BlancoPillado:2005fn} on the basis of this analysis was that all Minkowski supersymmetric vacua can have only positive masses, of flat directions, but never tachyons. 

An additional requirement we will impose on the class of models of our interest is that only models with strictly positive second derivatives are relevant for construction of de Sitter minima.
\begin{equation}
\phi_a \, V_{a \bar{b}}^ {Mink} \, {\bar \phi_{\bar b}}> 0 \ .
\end{equation}
It means that {\it the fermion mass matrix $m_{ab}$ should  not have zero modes}. In such case we can guarantee that Minkowski extremum is a local minimum. If the smallest, but non-vanishing, eigenvalue of the fermion mass is $m_\chi$ 
\begin{equation}
\phi_a \, V_{a \bar{b}}^ {Mink} \, {\bar \phi_{\bar b}} > m_\chi^2 \phi\bar \phi > 0 \ .
\end{equation}

In case of only one chiral multiplet with a non-vanishing fermion mass and a canonical \K\, metric we have reproduced the WZ feature shown in eq. \rf{WZ}. We will present below a class of models where Minkowski vacua are discrete and have no flat directions. A simplest example of such case is the KL model \cite{Kallosh:2004yh}.

\subsection{Downshift  to $W\neq 0$ and Supersymmetric AdS Minimum}
Here we slightly modify the superpotential so that it does not vanish at the extremum of the potential. In our notation this means that
 \be
{\it m}= e^{K\over 2} W \neq 0\, ,  \qquad {\it m}_a=0 \ ,
\ee
and 
\begin{equation}
(\mathcal{M}^2_{sc})^ {AdS}=\, \left ( \begin{array}{cc} V_{a \bar b}^ {AdS} & \, \,  V_{a  b}^ {AdS} \cr 
\cr
V_{\bar a  \bar b}^ {AdS} & \, \, \bar V_{\bar a  b}^ {AdS} \cr  \end{array} \right ) \ .
\label{massAdS}
\end{equation}
The change of the position of the minimum from $z_a^{Mink}$ to $z_a^{AdS}$ can be described in the general case. We would like to preserve the functional dependence of the \K\, potential and the superpotential of the moduli fields. We would like only to change the constant part of the superpotential by some small constant, $\Delta W$. Therefore first we will find the change of the position of the minima in AdS, versus the one in Minkowski. We require now that for generic values of $z$
\be
K^{AdS} (z, \bar z) = K^{Mink} (z, \bar z)\, , \qquad W^{AdS}(z) = W^{Mink}(z) + \Delta W \ .
\ee
The unbroken supersymmetry condition can be also given in the form
\be
m_a= e^{K\over 2} D_a W =0 \ .
\ee
One finds that the shift of the position of the minimum consistent with $m_a=0$ is given by the following formula
\be
\delta z^a = - (m_{ab})^{-1} K_b\, \Delta m +\dots
\ee
Here $\dots$ are terms smaller than the one we present,  and $\Delta m = e^{K\over 2} \Delta W \ll m_\chi$ where again we have neglected smaller terms. Thus we can see that if the mass of gravitino mass $\Delta m$ is parametrically small comparative to the smallest eigenvalue of the fermion matrix $m_{ab}$, the change of the position of the minimum is also small.

Now we can evaluate the change of the mass matrix in the supersymmetric AdS versus Minkowski. The mass formula at the point $z+\delta z$ is
\be
V_{a \bar b}^ {AdS} = {\it m}_{ac}g^{c\bar c} \,\bar {\it m}_{\bar c \bar b} - 2 g_{a \bar b}  {\it m} \bar {\it m} \ ,
\label{ads1}\ee
The first term here, as before is positive definite: the scalar masses are related to the square of the fermion masses. Therefore taking into account the change from $z$ to $z+\delta z$ in the first term is of no relevance since the total term is positive definite at any point, if the vacua remain discrete and do not acquire flat direction. But since the shift of the position of the minimum is parametrically small, our vacua remain discrete. The second term is negative, however, it is parametrically small comparative to the first term in \rf{ads1} and therefore will not change the positivity of the mass matrix.

The off-diagonal term in the mass matrix \rf{massAdS} was absent in Minkowski case, now it is present
\be
V_{a  b}^ {AdS} =  -    {\it m}_{ab}  \bar {\it m} \ .
\label{ads2}\ee
The condition that the supersymmetric AdS extremum remains a minimum, inherited from the Minkowski minimum, requires that the corrections to the mass matrix, which was positive definite in the Minkowski minimum, are small. To make all of these corrections irrelevant we need to require that the scale of the gravitino mass ${\it m}$ is much smaller that the scale of the fermion matter mass ${\it m}_{ab}$
\be
{\it m}_{ac}g^{c\bar c} \,\bar {\it m}_{\bar c \bar b} \gg 2 g_{a \bar b}  {\it m} \bar {\it m} \ .
\ee
Thus we need that the mass of gravitino is much smaller that the lightest eigenvalue of the matter fermion $|m_{\chi}|$ as shown in eq. \rf{chi}  
\be
m_{\chi} \gg {\it m}_{3/2} \ .
\ee
If we find a class of models satisfying this condition, we are guaranteed to find a class of AdS minima without tachyons.

\subsection{Uplift to dS Minimum}
Here we  add to our model a nilpotent multiplet $X$, describing also an uplifting anti-Dp-brane in string theory. 
\be
K^{dS}= K + K_{X, \bar X} X\bar X \ , \qquad W^{dS} = W+ \mu^2 X \ .
\ee
We now have additional terms to take into account like $
{\it m}_{X}= D_X {\it m},  {\it m}_{aX}= D_a D_X {\it m}, {\it m}_{abX}= D_a D_a D_X {\it m}$ etc. Thus for all superfields, including the nilpotent one, $\{a, X \}$,   will will use an index $I$, where   $I= \{a, X \}$. The value of the potential now is
\be
V_{dS} = e^K (|D_I W|^2  - 3 |W|^2) = |m_I|^2 - 3 |m|^2=    |F|^2 - 3m_{3/2}^2 >0\ .
\ee
Thus, the supersymmetry breaking terms $ |m_I|^2= |F|^2$ have to be at or above the scale of the gravitino.
We are looking now for the new extrema of the potential and the values of the mass matrix. Our goal is to describe  dS minima   with $\Lambda \geq  10^{-120}$. 
\be
V'=0\ , \qquad V=  |{\it m}_I|^2-3|{\it m}|^2 \geq  10^{-120} \ ,
\ee
The general  mass formula \rf{massM} in case that neither a superpotential nor its \K\, covariant derivative vanish can be given in the following form\footnote{We do not study the derivative of the potential in the direction of the nilpotent scalar $X$ since the scalar part of the nilpotent multiplet is not a fundamental scalar but is proportional to fermions.  }
\be
V_{a\bar b}^{dS}= {\it m}_{aI}g^{I\bar J} \,\bar {\it m}_{\bar J \, \bar b} - 2 g_{a \bar b}  {\it m} \bar {\it m}  + g_{a\bar b} m_I \bar m^{ I} - R_{a\bar b I \bar I} \bar m^I m^{\bar I} - {\it m}_{a} \bar {\it m}_{\bar b} \ ,
\label{dSmass}\ee
where  $R_{a\bar b I \bar I}$ is the moduli space curvature and $V_{a\bar b}^{dS}$ is the holomorphic-anti-holomorphic part of \rf{massM}. We also note that the new terms depending on $m_{aX}$ according to eq. \rf{fmass} have only terms proportional to $m_I$ since $\partial_a \partial_X W=0$ in our models.

The holomorphic-holomorphic mass formula in an uplifted dS model take the form
\be
V_{ab}^{dS} = -    {\it m}_{ab}  \bar {\it m}  + m_{abI} \bar m^I  \ .
\label{hh}\ee
The related mass formulae were established  in the form depending on holomorphic $W$ in \cite{Denef:2004ze} and used in the context of STU moduli stabilization   in \cite{Kallosh:2014oja}.

The mass formula at the dS extremum can be related to the one in the supersymmetric AdS before uplifting, taking into account that the shift of a position of the moduli at the minimum takes place, after the uplift terms are added. Note that the shift of the position of the dS minima versus the AdS minima is due to the uplifting terms and their dependence on moduli. For  $m_X= e^{K\over 2} D_X W$ 
\be
  |{\it m}_X|^2 = e^{K} \mu^4 \ .
\label{ineq1}\ee
We can see that parametrically small $\mu^2$ means that there is a small shift of a position of the minimum. We can make this statement  more specific by looking at the change in the potential due to uplifting terms and using real and imaginary fields
\be
z^a= z^a_1+ i z^a_2 \ .
\ee
The AdS part of the potential remains the same $V^{AdS}(z^a_1, z^a_2)$ the uplift term is given by 
\be
V^{\rm uplift}= \mu^4 e^K K_{X, \bar X} \ .
\ee
The new position of the dS minimum is at $z^a_\alpha$ with $\alpha=1,2$
\be
\partial_{z^a_\alpha} [  V^{AdS}+ V^{\rm uplift}]=0 \ .
\ee
The first term gives
\be
(\partial_{z^a_\alpha}  \partial_{z^b_\beta} V^{AdS} )\delta z^b_\beta \ ,
\ee
whereas the second on is
\be
\partial_{z^a_\alpha} V^{\rm uplift}= \mu^4 \partial_{z^a_\alpha} (e^K K_{X, \bar X}) \ .
\ee
We end up with an expression for the shift of the minimum
\be
\delta z^b_\beta = - (\partial_{z^a_\alpha}  \partial_{z^b_\beta} V^{AdS} )^{-1} \mu^4 \partial_{z^a_\alpha} (e^K K_{X, \bar X}) \ .
\ee
And since the scalar masses $(\partial_{z^a_\alpha}  \partial_{z^b_\beta} V^{AdS} )$ are parametrically greater than the scale of uplifting, we conclude that the shift of the minimum is small.

Using this formula for the shift of the moduli, one finds that in dS stage the value of supersymmetry breaking in the unconstrained chiral moduli directions is parametrically smaller than the supersymmetry breaking in the nilpotent direction, 
\be
|m_a|^2 \ll |m_X|^2  \ .
\ee 
This feature was established in the one-moduli case before in \cite{Linde:2011ja}. 

To make sure that our newly constructed  dS state is stable  we will require  that the supersymmetry breaking  is parametrically small, comparative with the scale of scalars and fermions in the chiral multiplets, in addition to our earlier requirement that the mass of gravitino is parametrically small, comparative with the scale of scalars and fermions in the chiral multiplets
 \be
 m_\chi ^2  \gg|F|^2= |m_I|^2 \ ,   \qquad  m_\chi ^2  \gg m_{3/2}^2 \ ,
\label{smaller}
\ee
and the potential is positive
\be
|F|^2 > 3 m_{3/2}^2 \ .
\ee
Since supersymmetry breaking in chiral directions is very small, $|{\it m}_a|^2  \ll  |{\it m}_X|^2   \ll  m_\chi ^2$ it means that
 supersymmetry breaking in the direction of the nilpotent superfield is of the order of gravitino mass for extremely small cosmological constant of the order $\Lambda \approx 10^{-120}$ or exceeds it for larger values of $\Lambda >0$.

The second derivative of the potential in \rf{dSmass} has a first term which is positive definite. It differs from its Minkowski value due to the shifts of the position of the minimum, however, by construction this term is positive definite. It is also strictly positive if our Minkowski vacua had no flat directions, since the shifts of the positions of the minima are small.
Note that all additional terms in \rf{dSmass} comparative to its positive definite first term ${\it m}_{ac}g^{c\bar c} \,\bar {\it m}_{\bar c \bar b}$ are small due to eq. \rf{smaller}.
We conclude that up to small terms involving $m$, $m_a$ and  $m_X$
\be
V_{a\bar b}^{dS} \approx V_{a \bar b}^ {Mink}  \ ,
\ee
and therefore
\be
\phi^a V_{a \bar b}^ {dS}  \bar \phi^{\bar b} \geq \phi\bar \phi \, m_\chi^2 >0 \ ,
\ee
where $m_\chi$ is the smallest eigenvalue of the matter fermion at the dS minimum.

The holomorphic-holomorphic mass formula in an uplifted dS model 
is shown in eq. \rf{hh}. The smallness of the first term $V_{ab}^{AdS}$ is due to a smallness of the down shift to AdS defined by a small value of $m$ versus matter fermions. In dS case \rf{hh} there is an additional term involving a third  holomorphic derivative of the covariantly holomorphic gravitino mass $m$ but it is multiplied by a small value of $\bar m_{\bar I} g^{I\bar I}= \bar m^I$. And since the susy breaking in all $a$ directions is much smaller that in the nilpotent direction, the only term to consider is $m_{abX} \bar m^X$. In models where the superpotential has a simple dependence on $X$ of the form $\mu^2 X$ we find that  $m_{abX}=0$. The term $m_{abc} \bar m^c$ has an additional small factor since $|{\it m}_a|^2  \ll  |{\it m}_X|^2$.

Thus, the off-diagonal terms  $V_{ab}^{dS}$
in the mass matrix \rf{massM} are negligibly small, 
and the mass matrix in dS vacuum remains  positive definite, without flat directions, as inherited from the Minkowski vacuum.

As we have shown above, the explicit choices of $K$ and $W$, particularly motivated by string theory like the multifield KL models, are available, in which the mass production of dS minima is guaranteed by an evaluation of the  shift in the position of the vacuum and the mass formula studied here, where the conditions for its positivity are specified.

For unspecified $K$ and $W$ it is useful to have the exact  formulas for the second derivatives of the potential in eqs. \rf{dSmass}, \rf{hh}.  The values of the moduli space metric $g_{a\bar b}$,   the moduli space curvature $R_{a\bar b I \bar I}$ and the third holomorphic derivative of the superpotential $m_{abI}$ may indicate what is the required level of smallness of $m_{3/2}$ to guarantee that all terms in eqs. \rf{dSmass}, \rf{hh} remain small comparative to the first term in eq.  \rf{dSmass} which is always positive.

\section{String theory embedding of  the new 4d supergravity models}\label{emb}
The main focus of this paper was on a new construction of dS minima in 4d supergravity inspired by type IIA string theory.  Here we would like to address  some issues of the string theory embedding of our new supergravity models. We note that in this paper
the only difference with the setup in \cite{Cribiori:2019bfx} is that the non-perturbative part of $W$ has two (or more) exponential terms for each moduli, whereas in  \cite{Cribiori:2019bfx} a single exponent in $W$ for each moduli was used.

 A discussion of the relation of the models developed in \cite{Cribiori:2019bfx} to string theory is contained in  section 2.2 of \cite{Cribiori:2019bfx}, with the title `Satisfying stringy requirements.' For example, it is pointed out there that the only non-trivial Bianchi identity is the tadpole condition for the D6-brane charges, since all other RR as well as NS-NS fluxes are absent, we have no Romans mass and also the 2- and 4-fluxes are absent, and $H_3=0$. The relevant equation is
$
\int dF_2 -F_0 H = -2 N_{O6} + N_{D6} - N_{\overline {D6}}\ ,
$
which has to be satisfied for each three-cycle independently. It is explained in  \cite{Cribiori:2019bfx} how it is possible to satisfy this equation. In the current paper the only difference with  \cite{Cribiori:2019bfx} is only  in the non-perturbative exponents, here we have two of them for each modulus, and this  does not affect the tadpole condition. Therefore  we believe that there are no problems with the tadpole condition in the current models. The details will be worked out in future studies.

Other interesting points concerning the relation of these models to string theory are with regard to  10d classical background on which the Type IIA string theory is compactified (before the introduction of instantons and anti-D6 branes). 
Type IIB GKP models \cite{Giddings:2001yu}  involve the  4d superpotential  $W=\int G_3\wedge \Omega$ with
imaginary self-dual (ISD) 3-flux $G_3=F_3-\tau H_3 $ and $G_{(0,3)}=0$. Supersymmetric solutions  require $G_3$ to be only of the type (2,1). In GKP the  moduli stabilization is studied in the context of the  compact version of Klebanov-Strasssler throat.
In supersymmetric Minkowski solutions $DW=W = 0$ in the vacuum, and it was   shown by GKP that it is possible to stabilize the dilaton-axion field $\tau$. To compare the GKP setup with the type IIA models (before the non-perturbative terms are added to $W$) we notice that we only have $F_6$, which  breaks supersymmetry. For example, to have unbroken Minkowski supersymmetry one needs $F_0= F_6$ as well as other conditions. But we do not have the Romans mass, $F_0=0$, therefore in our type IIA models (before the non-perturbative terms in $W$ are included) Minkowski supersymmetry is always broken.

To compare with KKLT, we first notice that Minkowski supersymmetry is broken in KKLT (before the non-perturbative terms in $W$ are included), since in KKLT, as opposite to GKP, $G_{(0,3)}\neq 0$, which results in a constant term $W_0$. Once the exponent is added, one finds  AdS  supersymmetric vacua.
In our models the constant term $W_0$ in the superpotential originates from $F_6$ flux. Once we add non-perturbative terms, if it is a single exponent, we find  supersymmetric AdS minima, as in KKLT.  These were described in  \cite{Cribiori:2019bfx}, and they were uplifted to dS minima using anti-D6 branes, in a complete analogy with KKLT uplift using anti-D3 branes. Thus in  \cite{Cribiori:2019bfx} we had a type IIA solution of the KKLT-type.
In the current paper, we added a second non-perturbative exponent, which  enabled us to get the Minkowski  supersymmetric vacua, followed by supersymmetric AdS vacua, which were uplifted to dS vacua.

It is also interesting that in type IIB in GKP there are non-vanishing $F_3$ and $H_3$ fluxes. Our type IIA models have a six-flux $F_6$ which is dual to $F_3$ in type IIB, but the $F_2$ and $F_4$ are absent. Moreover, the NS-NS 3-flux $H_3$ is also absent in our type IIA models. In particular the integer $K$ associated with the integral over a 3-cycle over $H_3$, which plays a prominent role in Klebanov-Strassler throat, is vanishing in our type IIA models. Therefore type IIA appears to be simpler than type IIB.

 The main problem in the past was to uplift type IIA  AdS vacua to dS, various no-go theorems are known, see e.g.  \cite{Kallosh:2006fm, Hertzberg:2007wc, Haque:2008jz, Flauger:2008ad, Danielsson:2009ff, Wrase:2010ew, Shiu:2011zt, Junghans:2016uvg, Andriot:2016xvq, Junghans:2016abx, Andriot:2017jhf}. The corresponding  uplifting mechanism via the anti-D6-brane was discovered relatively recently, in  \cite{Kallosh:2018nrk}, whereas the uplifting role of the anti-D3-brane was known long time ago since the  KKLT construction in type IIB theory. Another reason why in type IIA models we have now presented dS vacua, overcoming the no-go theorems, is that we have used the non-perturbative exponents in the superpotential for all moduli directions, motivated by the U-duality of the non-perturbative string theory. This fact was also discussed in the previous paper  \cite{Cribiori:2019bfx} where  single KKLT type exponents were used, whereas in  the current paper we use racetrack superpotentials with two or more exponents. A further investigation of  the string theory embedding of the mechanism of mass production of dS vacua in supergravity, with and without non-perturbative terms,  is given in our next papers  \cite{Cribiori:2019drf,Cribiori:2019hrb}.

\section{Conclusions}

Mass production is the manufacture of large quantities of standardized products,   often using assembly lines. An assembly line is a manufacturing process  in which parts  are added as the semi-finished assembly moves from workstation to workstation where the parts are added in sequence until the final assembly is produced. 

In our case  the final products are the  metastable de  Sitter vacua,
and we have three stages in the production. The first stage involves  construction of  stable Minkowski vacua without flat directions in models with multiple scalars.  The crucial observation made in the present paper is that if one takes a sum of $n$  superpotentials $W_i (T_{i})$ for each of the moduli $T_{i}$, the problem of finding a stable supersymmetric minimum for all of these moduli becomes divided into $n$ independent problems for each of these moduli. To find a supersymmetric Minkowski vacuum at any point $T^{0}_{i}$ one should identify $n$  superpotentials $W_{i}(T_{i})$ which have a vanishing first derivative at $T^{0}_{i}$, take a sum of all  of these superpotentials, and subtract from them the sum of all of these superpotentials at $T^{0}_{i}$. This simple procedure automatically produces a theory with a stable supersymmetric Minkowski vacuum at $T^{0}_{i}$, in absence of  flat directions.

A simple example of string theory motivated supergravity models of this kind involves the racetrack KL-type  superpotentials. The problem of construction of stable Minkowski vacua in a theory of this type with a single modulus was solved in  \cite{Kallosh:2004yh}. The procedure outlined above automatically generalizes this solution for any number of moduli.
Such supersymmetric Minkowski vacua are possible in superstring theory, and they are  stable  \cite{BlancoPillado:2005fn}. 

However, this is not a finished product yet, for two reasons. First of all, we do not live in a supersymmetric world, so we need to break SUSY. But we should not break it too much because it can violate the stability of the vacuum that we constructed. Secondly, we would like to find a mechanism generating a dS vacuum with a tiny cosmological constant $\Lambda_{dS}\approx 10^{-120}$, in Planck units. Both of these goals can be reached at the next two stages of the procedure.

The second stage describes how to preserve the stability of the Minkowski vacuum when downshifting it  to  a supersymmetric AdS vacuum with a negative cosmological constant of the order of the gravitino mass squared, $\Lambda_{AdS} = V_{AdS}=- 3m_{3/2}^2$. 

The third stage leads to the final product, a metastable  dS vacuum, when one adds an uplifting contribution of the  $\overline {D6}$-branes  \cite{Kallosh:2018nrk,Cribiori:2019bfx}, corresponding to a nilpotent multiplet in supergravity. Here  we assume that   the total SUSY breaking parameter $F^2 = e^K (|D_a W|^2 + |D_X W|^2)\approx e^K  |D_X W|^2 $ overcompensates for the effect of the downshift and in this way many different stable dS vacua can be produced. In those cases where the uplift only very slightly  overcompensates  the  downshift, we find dS vacua with  very small values of the cosmological constant. This is a part of the  standard string theory landscape construction, which is necessary to account for the  tiny positive value of the cosmological constant $\Lambda_{dS}\approx 10^{-120}$.  The resulting dS state inherits the stability property of its (grand)parent supersymmetric Minkowski vacuum if the resulting supersymmetry breaking in dS vacuum is sufficiently small. 

Examples of such models involving multiple KL models  \cite{Kallosh:2004yh} consistent with type IIA string theory inspired supergravity are presented in the paper, see section 3 and appendix~A, and STU model in section 4. More general explicit examples will appear in the future. In discussing the relation between the parameters of our multiple KL models with type IIA  string theory one has to keep in mind that it is easy to change the parameters via rescaling without affecting stability. In particular, in the context of the KL-related scenario developed in this paper, the superpotentials for each field $T_{i}$ are described by 4 parameters $a_{i}$, $b_{i}$, $A_{i}$ and $B_{i}$. For each such set, one can take  $a_{i} >b_{i}$ without any loss of generality. Then a stable supersymmetric vacuum can be constructed for any values of these parameters with $a_{i} A_{i} > b_{i} B_{i}$, which is not much of a restriction.

Our results obtained in section 5 show that the  conclusions about stability of dS vacua at stage three of production are valid for more general \K\ potentials and superpotentials. The conditions for this require that 1) the supersymmetric Minkowski vacuum does not have flat directions,  2) mass of gravitino at stage two and susy breaking at stage three are parametrically small, comparative to the scale of scalars stabilized at the first stage of production, in Minkowski vacuum.  If these conditions are satisfied, there is no further restriction on the functional form of $K$ and $W$.

 Moreover, instead of adding a small constant $\Delta W$ to the original superpotential $W$ during the downshift stage, one can add a small function of all moduli fields $\Delta W(T_{i})$ depending on all moduli fields. If this function is sufficiently small,   it can 
shift the minimum of the potential to AdS, but it does not affect the positive definiteness of the mass matrix, which is also preserved by the subsequent small uplift. Thus one can significantly generalize the mechanism of dS vacua production introduced in our paper, while preserving its qualitative features and consequences. 

It would be interesting to find more general explicit models of this kind, which lead to stable dS vacua in supergravity, and can be eventually associated with various versions of string theory. We discussed the issues of embedding our models into string theory in section \ref{emb}.
A more detailed investigation of  the string theory embedding of our proposal on a mass production of dS vacua in supergravity is contained in our next papers  \cite{Cribiori:2019drf,Cribiori:2019hrb}.  We study there a  compactification on specific CY$_3$ manifolds, like K3 fibration models, a CICY model, and a multi-hole Swiss cheese model \cite{Cribiori:2019drf}, as well as a compactification on  $G_2$  in M-theory \cite{Cribiori:2019hrb}. In these papers we also discuss  the issue of tadpole conditions and other global string theory related constraints. We expect more studies of this kind in  the future.

 \noindent{\bf {Acknowledgments:}} We are grateful to J. J. Carrasco, N. Cribiori, S. Kachru, C. Roupec, E. Silverstein, and T. Wrase for stimulating discussions and important comments.
 This work is supported by SITP, by the US National Science Foundation grant PHY-1720397, by the Simons Foundation Origins of the Universe program (Modern Inflationary Cosmology collaboration), and by the Simons Fellowship in Theoretical Physics.
 
 \appendix 

\section{ Some illustrative examples}

In this section we will illustrate our general approach by a numerical analysis of the simplified STU model describing two fields  $U$ and $T$  and the nilpotent field $X$ with 
\be\label{TU1}
W = W_{0}+\Delta W+ A_{u} e^{-a_{u}U} -  B_{u} e^{-b_{u}U} + A_{s} e^{-a_{s}S} -  B_{s} e^{-b_{s}S} + \mu^2\, X \ ,
\ee
\be\label{TU2} 
K = - 3\ln(U+\overline{U}) - 4\ln(S+\overline{S})     +  \frac{X \bar{X}}{(U+\overline{U})^{3}(S+\overline{S})} . 
\ee

\subsection{A very small downshift and uplift}
As an example, we will look for a supersymmetric Minkowski vacuum with $\Delta W = 0$ and $\mu^2 = 0$ at  $U = u_{0} = 5$ and $S = s_{0}= 1$. We will take the following set of parameters: $a_{u} = 1$, $b_{u} = 1.2$, $A_{u} = 10$, $a_{s} = 2$, $b_{s} = 3$, $A_{s} = 1$.  The parameters $B_{s} ={2 e\over 3}$ and   $B_{u} =22.6523$ are found using \rf{mingen} and the requirement that the minimum   is at $u_{0} = 5$ and $ s_{0}= 1$.  

\begin{figure}[h!]
\begin{center}
\includegraphics[scale=0.55]{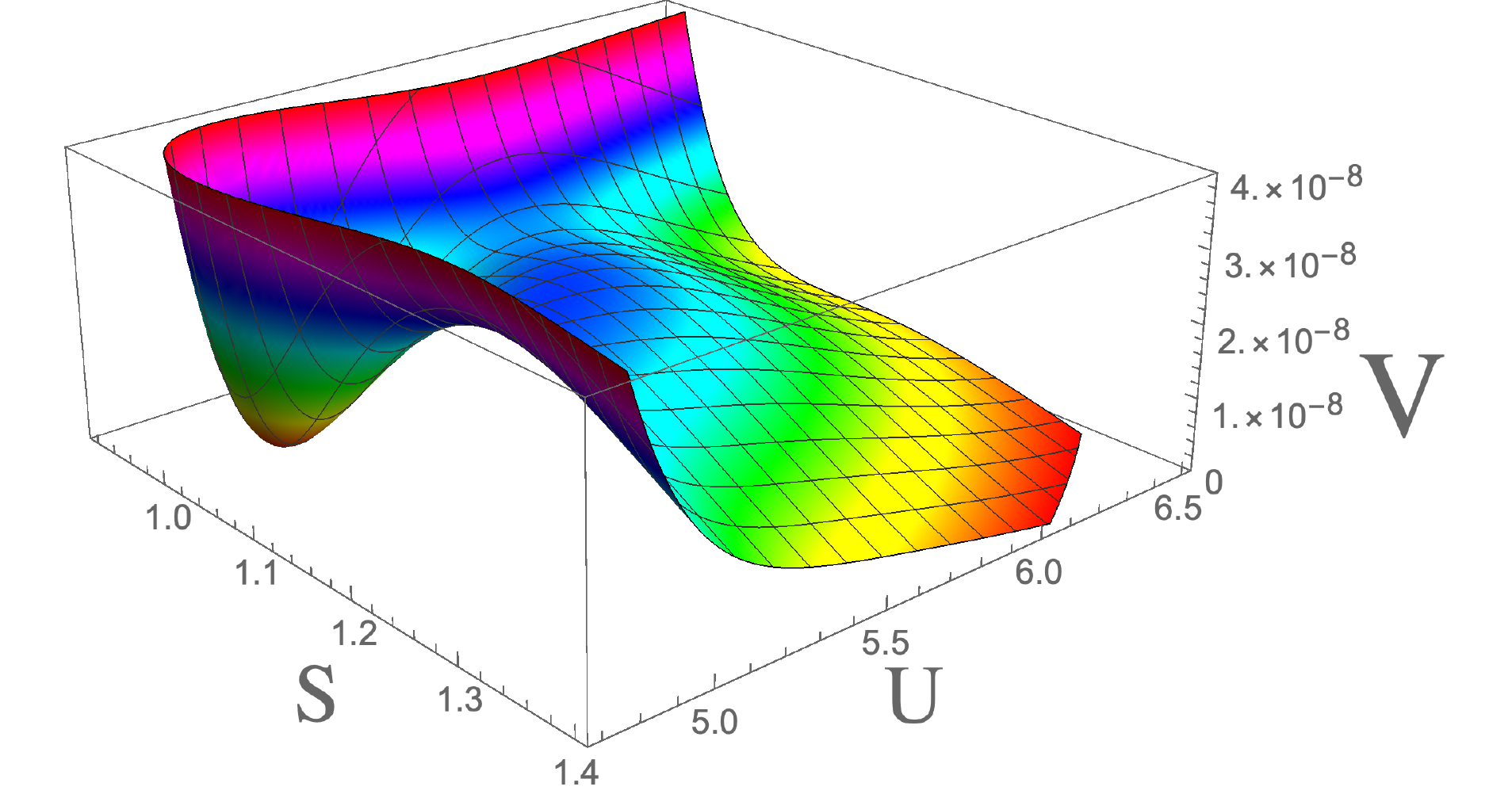}
\end{center}
\vskip -0.5cm 
\caption{\footnotesize  The potential of the model \rf{TU1}, \rf{TU2} with respect to the real parts of the field $S$ and $U$, for $\Delta W = 0$ and $\mu^2 = 0$ in the Planck mass units. The supersymmetric Minkowski minimum with $V = 0$   is at $S = 1$ and $U = 5$.}
\label{f1}
\end{figure}

With these parameters, using \rf{w0}, one finds the value  $W_{0}$ required for the existence of the supersymmetric Minkowski minimum for $\Delta W = 0$ and $\mu^2 =0$: \, $W_{0}= -0.0563417$. The resulting potential $V(T,U)$ is shown in Fig. \ref{f1}.

Now we will study the same potential by including the contribution 
\be
\Delta W = 10^{{-5}} \ ,
\ee
 and, in the next step, including the term   $\mu^2\, X$ with 
\be
\mu^2 \gtrsim 3.873\times 10^{-7} \ ,
\ee 
which is just sufficient to compensate the downshifting the potential  by $\Delta W = 10^{{-5}}$, and by making the value of the potential at its minimum positive.  To make these small changes visible, we will zoom at a very small vicinity of the minimum of the potential shown in Fig. \ref{f1}, and then show how it changes when we introduce $\Delta W = 10^{{-5}}$, and when we subsequently uplift the potential by including the contribution $\mu^2 X$ of the nilpotent field $X$. 

\begin{figure}[t!]
\begin{center}
\begin{center}
\includegraphics[scale=0.46]{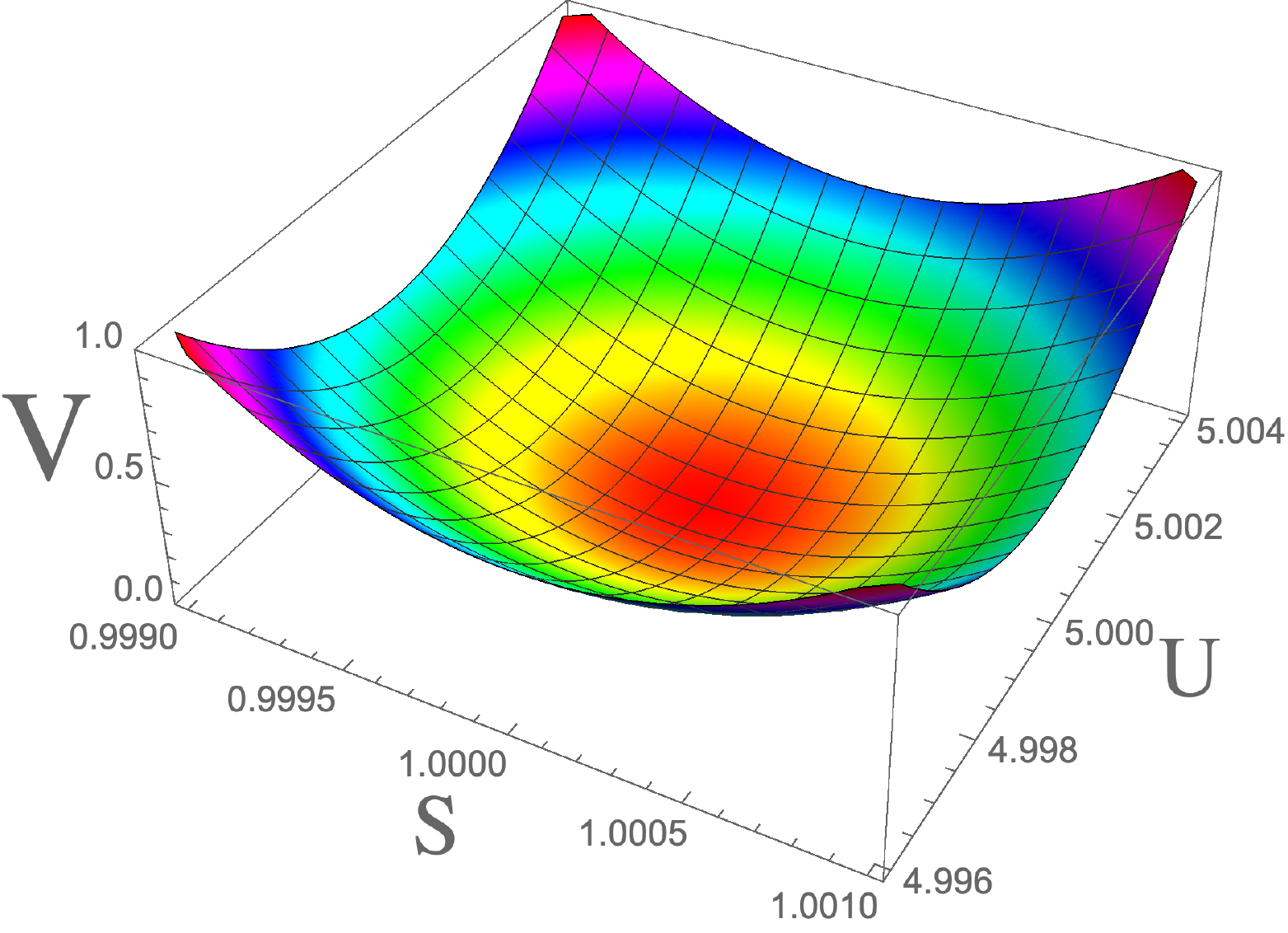}  
\includegraphics[scale=0.46]{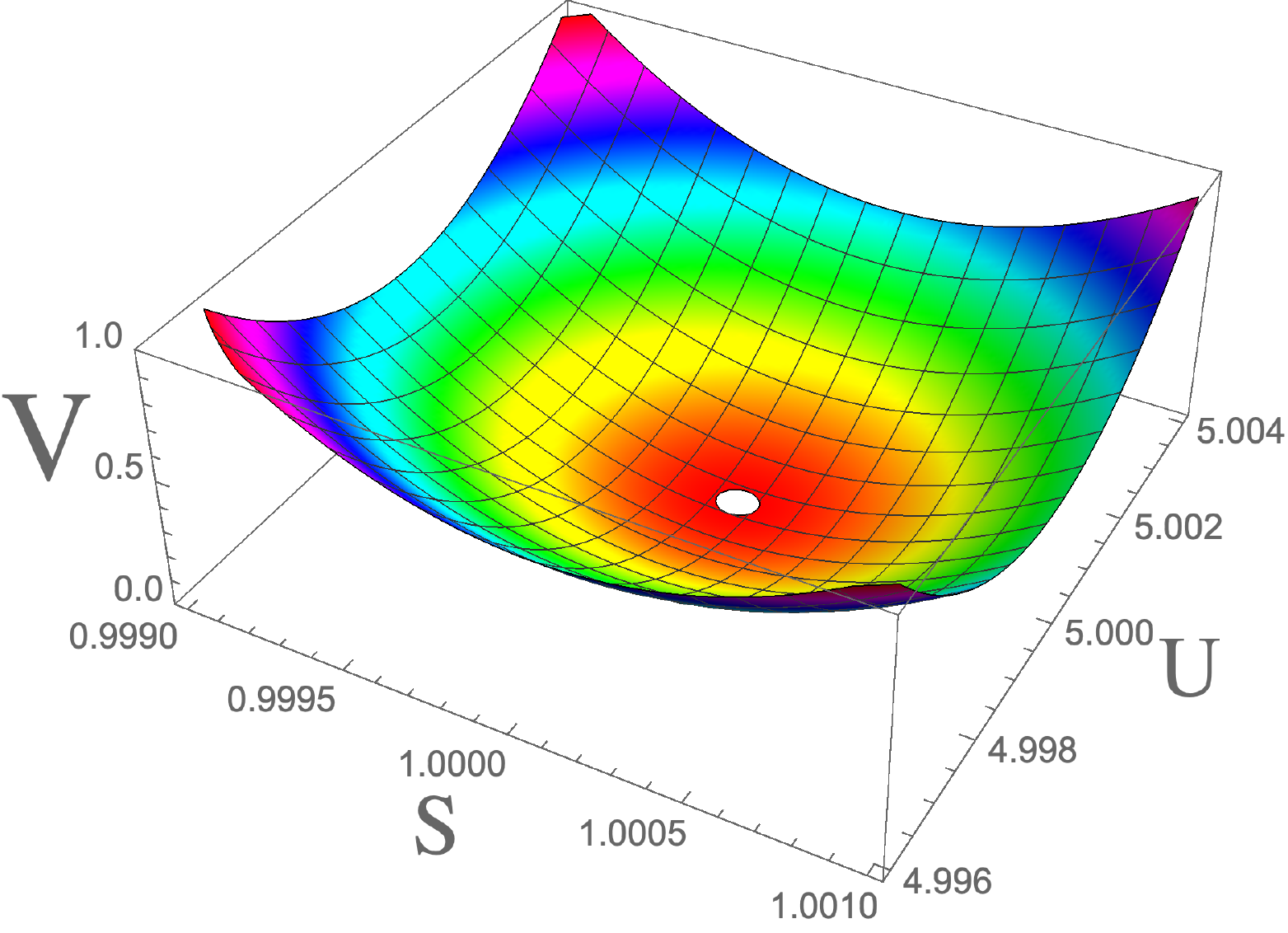}  
\includegraphics[scale=0.46]{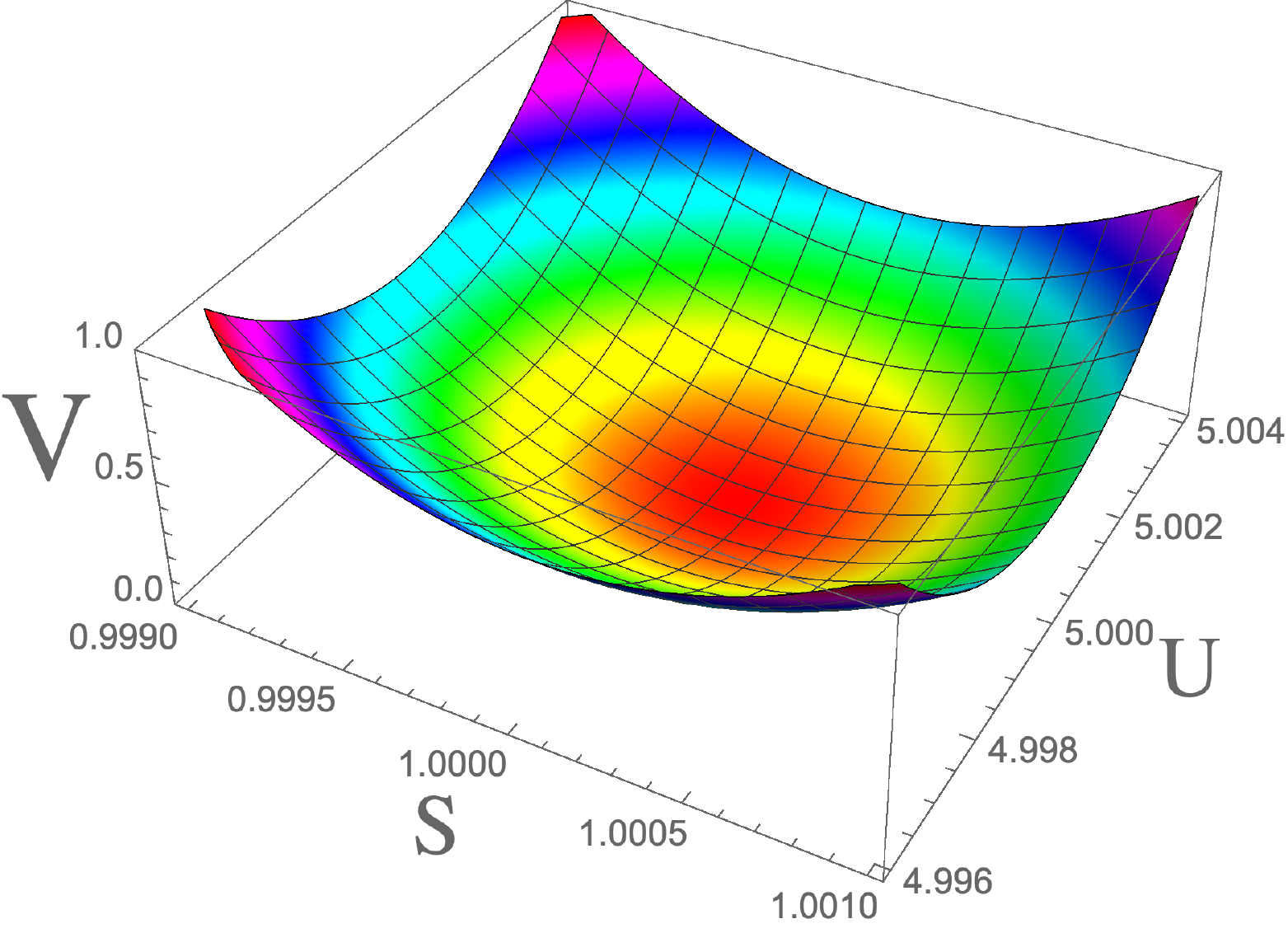} 
\end{center}
\end{center}
\vskip -0.5cm 
\caption{\footnotesize  The potential of the model \rf{TU1}, \rf{TU2}, Fig. \ref{f1} with respect to the real parts of the field $S$ and $U$, in a small vicinity of its minimum. It is shown in units $10^{{-11}} M_{p}^{4}$ , which is 3000 times smaller than the height of the potential shown in Fig. \ref{f1}. The upper left figure shows the potential for $\Delta W = 0$ and $\mu^2 = 0$, with the supersymmetric minimum at $V = 0$, $S = 1$ and $U = 5$. The upper right figure shows the potential for $\Delta W = 10^{{-5}}$ and $\mu^2 = 0$. It has a minimum at $V<0$. Instead of showing it, we cut the potential at $V = 0$. Thus the white hole in the ground clearly shows that the potential below it is negative. The lower figure shows the stable uplifted dS vacuum with $V > 0$, which requires including the contribution of the nilpotent field $X$ with $\mu^2 \gtrsim 3.873\times 10^{-7}$.}
\label{f2}
\end{figure}

 As one can see, everything goes as expected: after a small downshift and uplift we have a stable dS minimum with respect to the real parts of the fields $U$ and $S$. One can check that the same is true for the potential of the complex components of these fields.

It is instructive to compare the mass matrix of all (canonically normalized)  fields in the supersymmetric Minkowski state and in the uplifted dS state. In the Minkowski state, the masses of the real  components of the fields $U$ and $S$ are given by 
\be
 m_{u} =    3.5512 \times 10^{{-3}}\ , \qquad  m_{s}= 2.1398 \times 10^{{-3}} \ ,
\ee
and the masses of the imaginary components have the same values.
After the uplift, the mass matrix remains positively definite, but requires diagonalization. The eigenstates for the real components of the fields $U$ and $S$  have masses 
\be
 m_{1} =    3.5488 \times 10^{{-3}}\ , \qquad  m_{2}= 2.1386 \times 10^{{-3}} \ ,
\ee
The masses of the imaginary components of the fields $U$ and $S$ are
\be
m_{3}=3.5489 \times 10^{{-3}}\ , \qquad  m_{4} = 2.1387 \times 10^{{-3}} \ .
\ee
In other words, the moduli masses do not change  much during the downshift and the uplift, and the potential also does not change much, see Fig. \ref{f2}. Thus dS state is stable. Gravitino mass in this state is $m_{3/2} = 1.26491 \times 10^{-6}$, which is 3 orders of magnitude smaller than other masses. Stability of the dS vacua for small values of $m_{3/2}$ is indeed expected on the basis of the investigation in section \ref{mass}.

\subsection{A significant downshift and uplift}

To check how stable this result with respect to the more significant downshift and uplift, we added $\Delta W = - 0.0036583$, to bring $W_{0}$ from  $-0.0563417$ to $-0.06$. Thus  change of $W$ is almost 3 orders of magnitude greater than $\Delta W = - 0.00001$ in the previous example. This change requires uplift with $\mu^2 \sim 1.44 \times 10^{-4}$. The resulting modification of the potential is quite significant, and yet our procedure yields a stable dS vacuum.  The mass matrix of all fields in the dS state is positively definite. In particular, masses of the real components of the fields $U$ and $S$ after the mass matrix diagonalization are $2.6712 \times 10^{-3}$ and  $1.658\times 10^{{-3}}$. 

The gravitino mass is $m_{3/2} = 4.835 \times 10^{{-4}}$. It  is only about 4 times smaller than the lightest moduli mass, the downshift and uplift significantly modify the mass matrix, but the dS state remains stable. 

This model  allows lots of freedom in the choice of the parameters, which can change the final results, see section \ref{baseKL}. For example, one can simultaneously rescale $A_{u}$ and $B_{u}$ by some factor $C$ without changing the position of the supersymmetric Minkowski vacuum. However, this change would increase the mass of the field $U$ by a factor of $C$, see \rf{mmatr} and \rf{dder}. A simultaneous rescaling of $A_{s}$ and $B_{s}$ by the same factor  $C$ would similarly increase the mass of the field $S$, and significantly strengthen the vacuum stabilization.  Meanwhile a simultaneous decrease of $a_{u}$ and $b_{u}$  would increase  $u_{0}$, and a simultaneous decrease of $a_{s}$ and $b_{s}$  would increase  $s_{0}$  \cite{Kallosh:2011qk}. This shows that the scenario discussed above is quite flexible. 

\subsection{A large uplift without downshift}

In the previous subsections we studied the downshift and uplift to a dS state with an extremely small value of the cosmological constant $V_{dS} \sim 10^{{-120}}$. However, if one is interested in the general structure of string theory landscape after uplift from the supersymmetric Minkowski vacuum, one may study a possibility of a very large uplift, limited only by a possible instability of the dS state after the uplift. Therefore in the example studied below we ignore the stage of the downshift to AdS, directly uplift the stable Minkowski vacuum shown in Fig. \ref{f1}, and verify its stability.

\begin{figure}[h!]
\begin{center}
\includegraphics[scale=0.55]{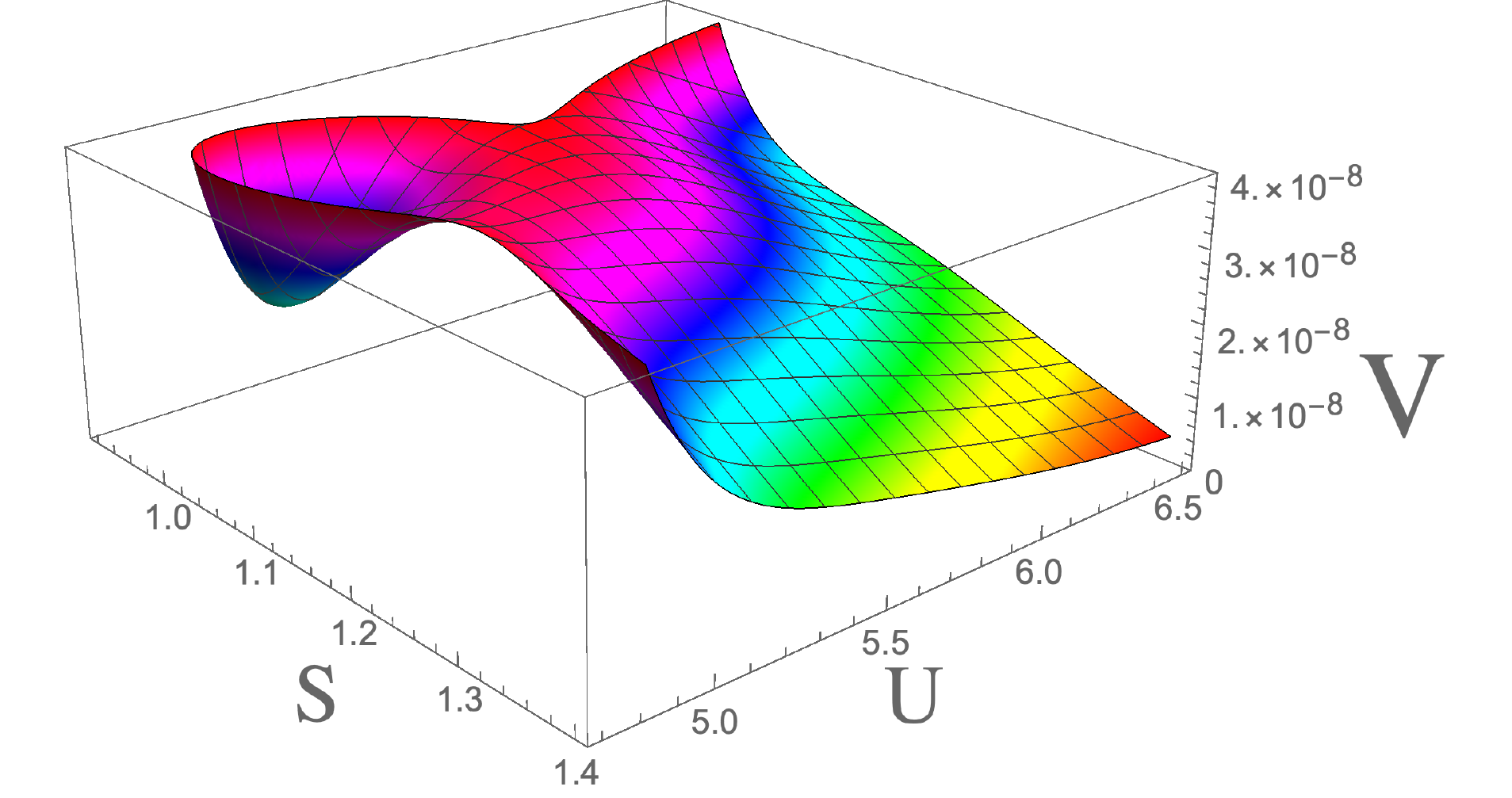}
\end{center}
\vskip -0.5cm 
\caption{\footnotesize  The potential of the model \rf{TU1}, \rf{TU2} for $\Delta W = 0$ and $\mu^2 = 4\times 10^{-4}$ in the Planck mass units. This figure illustrates the possibility to obtain a stable dS vacuum with large $V_{dS}$ by a direct uplift of the supersymmetric Minkowski vacuum.  The value of $V_{dS}$ in this vacuum is  $2\times 10^{-8}$, which is comparable with the height of the barrier stabilizing the original  Minkowski vacuum shown in Fig. 1.}
\label{f3}
\end{figure}

The height of the potential barrier stabilizing the Minkowski vacuum in the model shown in Fig. \ref{f1} is  $2\times 10^{-8}$, in Planck density units. In Fig. \ref{f3} we show that by taking the uplift parameter $\mu^2 = 4\times 10^{-4}$ one can uplift the Minkowski minimum to a stable dS minimum with $V_{dS} = 2\times 10^{-8}$, i.e. close to the height of the barrier shown in Fig. \ref{f1}. We also verified that this minimum is stable with respect to the imaginary components of the fields $S$ and $U$. As we mentioned in this paper, in the class of models we consider here, the mixed real-imaginary mass terms vanish. Thus the uplifted dS vacuum shown in  Fig. \ref{f3} is (meta)stable with respect to all fields.
By changing parameters of the model, one can produce stable uplifted dS states with even much greater vacuum energy density.

\bibliographystyle{JHEP}
\bibliography{lindekalloshrefs}
\end{document}